\documentclass[preprint,12pt]{elsarticle}

\usepackage{amssymb}
\usepackage{lineno}

\usepackage{caption}

\usepackage[perpage]{footmisc}

\journal{Nucl.\ Instrum.\ and Methods A}

\begin{document}

\begin{frontmatter}



\title{A Radiation-Hard Dual-Channel 12-bit 40 MS/s  ADC Prototype for the ATLAS Liquid Argon Calorimeter Readout Electronics Upgrade at the CERN LHC}


\renewcommand{\thefootnote}{$\dagger$} 

\author{J.~Kuppambatti$^a$,\footnote{Currently at Seamless Devices Inc., San Jose, CA} J.~Ban$^b$, \renewcommand{\thefootnote}{$\ddagger$} 
T.~Andeen$^b$,\footnote{Currently at the University of Texas at Austin} R.~Brown$^b$, R.~Carbone$^b$, P.~Kinget$^{a}$, G.~Brooijmans$^{b}$, W.~Sippach$^{b}$     }

\address{
$^a$Columbia University, Dept. of Electrical Engineering\\
  New York, New York, USA\\
  $^b$Columbia University, Nevis Laboratories\\
  Irvington, New York, USA\\
}

\begin{abstract}
The readout electronics upgrade for the ATLAS Liquid Argon Calorimeters at the CERN Large Hadron Collider requires a radiation-hard ADC. The design of a radiation-hard dual-channel 12-bit 40 MS/s pipeline ADC for this use is presented. The design consists of two pipeline A/D channels each with four Multiplying Digital-to-Analog Converters followed by 8-bit Successive-Approximation-Register analog-to-digital converters. The custom design, fabricated in a commercial 130~nm CMOS process, shows a performance of 67.9~dB~SNDR at 10~MHz for a single channel at 40~MS/s, with a latency of 87.5~ns (to first bit read out), while its total power consumption is 50~mW/channel. The chip uses two power supply voltages: 1.2 and 2.5~V. The sensitivity to single event effects during irradiation is measured and determined to meet the system requirements.  
\end{abstract}


\end{frontmatter}

%\linenumbers

%\section{}
%\label{}

\section{Introduction} \label{sec:Intro}

This article describes the design and performance of a radiation-hard dual channel 40 MS/s pipeline ADC prototype. A future version of this ADC is intended for the upgraded trigger electronics in the ATLAS Liquid Argon (LAr) Calorimeter readout. The current prototype is used to establish the analog performance of the four 1.5-bit pipeline stages and the 8-bit Successive-Approximation-Register (SAR) ADC stage, and to study the radiation tolerance of the ADC design. This work is an extension of the ADC design in \cite{nevis10}. There, the performance of the four 1.5-bit pipeline stages was presented, verifying the precision of the design as well of as the radiation hardness of the pipeline stages. This article introduces the SAR ADC stage, its integration with the pipeline stages, and a measurement of the sensitivity of the ADC to single event upsets (SEU) during irradiation.  

The Large Hadron Collider (LHC) \cite{lhc} at CERN is a proton-proton collider designed to operate at center-of-mass energies of 7--14 TeV. The LHC began circulating 6.5 TeV beams in 2015 and is expected to operate at approximately this energy for the rest of the decade. The ATLAS detector~\cite{jinst} is a multi-purpose apparatus built to observe the decays of particles produced by the high energy collisions at the LHC. The position and energy of electrons and photons from these decays are measured by the electromagnetic calorimeter system. This system consists of three calorimeters (a barrel and two end-caps in separate cryostats). These sampling calorimeters are read out by an analog and digital electronics chain. To precisely record the large dynamic range signals produced by the calorimeters and to limit noise, part of the electronic chain (the front-end) is located on the detector~\cite{FE}. 

Providing data to the on-line event selection (trigger system) is an important function of the LAr calorimeters. Electron and photon triggers are critical to the ATLAS physics program. The acceptance rate of these triggers will be affected by the planned LHC upgrades. Upgrades to the LHC are scheduled to occur in two stages, starting with Phase 1 in 2019-2020 and then for the High Luminosity LHC (HL-LHC) in 2024-2026~\cite{hllhcscoping}. These upgrades will increase the instantaneous luminosities from the nominal value ($10^{34}~\mathrm{cm}^{-2}\mathrm{ s}^{-1}$) to three times that for the data collected in 2021-2023 (after Phase 1)~\cite{phase1tdr}. For electron and photon triggers, which rely on the LAr calorimeter data, higher instantaneous luminosities would require increased energy thresholds. To avoid raising the thresholds additional data can be provided by LAr calorimeters to the trigger system. This allows the use of more efficient algorithms to reduce background events~\cite{phase1tdr}. 

\begin{figure}[h]
\includegraphics[width=32pc]{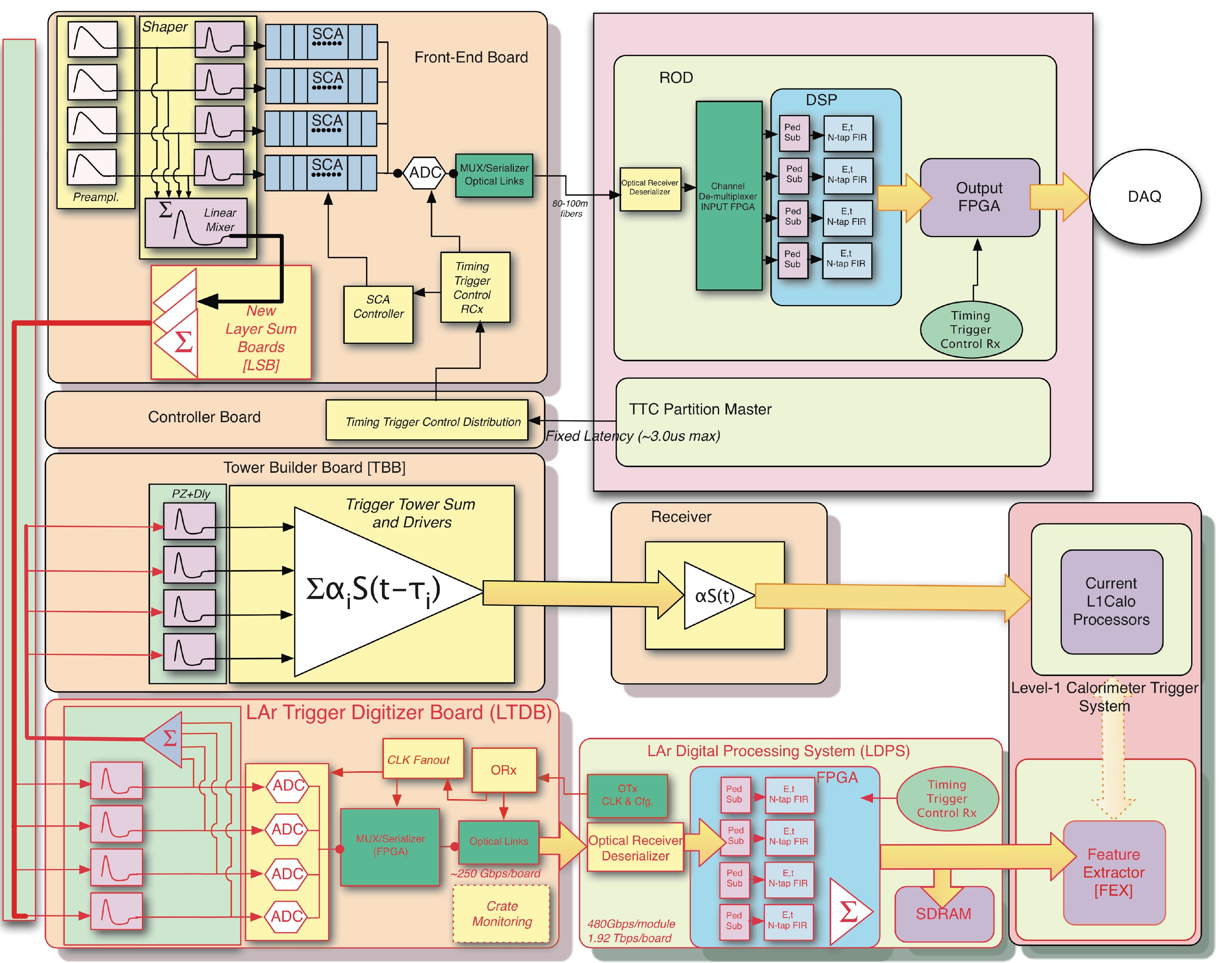}
\caption{\label{phase1block}Block diagram for the proposed ATLAS Phase-I electronics upgrade~\cite{phase1tdr}. The analog signals enter the system on the upper left, and are passed to the trigger system and DAQ system. The position of ADC is indicated in the lower left box (LTDB), digitizing the signals for the trigger system.}
\end{figure}

Currently, the existing front-end electronics limit the granularity, bandwidth and latency of the data transmitted to the lowest level of the trigger system. With the addition of new interface electronics between the LAr calorimeters and trigger system higher granularity data can by provided at the bunch crossing rate of 40~MHz. This will take the form of a LAr Trigger Digitizer Board (LTDB), illustrated on the bottom left of Figure \ref{phase1block} and to be installed as part of the Phase 1 upgrade. The design and strategy of the LAr calorimeter trigger electronics upgrade is described in greater detail in~\cite{phase1tdr}. A suitable ADC is an important requirement of this upgrade. Though to be installed as part of the Phase 1 upgrade, this ADC is designed to operate throughout the HL-LHC run. 

\section{ADC Architecture}

\subsection{ADC Specifications} \label{sec:sys_specs}
This ADC is designed to digitize the analog signals from the LAr calorimeters before they are transmitted to the trigger system. The raw calorimeter signal is a triangular pulse about 500~ns long. This signal is passed through a preamplifier and is then subject to a fast bipolar $CR-(RC)^{2}$ shaping function with $\tau$~=~RC~=~20~ns. In the precision readout of the detector these shaped signals are further amplified with a gain of 1, 10 or 100 to reduce the dynamic range requirements. However, for the trigger readout four channels are summed before shaping and no gain is applied. In the upgraded system these shaped signals will be sent to new Layer Sum Boards~\cite{phase1tdr}, which will combine the output of the shapers for each of the four calorimeter layers. In each layer either four or eight channels will be combined into one signal to be digitized. This will provide higher granularity readout when compared to the existing system of ``trigger towers.'' The ``trigger towers''  combine additional cells laterally, as well as the four layers, into a single tower. 

The ADC for Phase 1 must meet the following specifications:
\begin{itemize}
\item The calorimeter signals must be continuously sampled and digitized at (minimally) a frequency of 40 MHz.
\item The ADC power must be less than 145~mW.
\item The latency must be less than 200~ns. 
\item It must be radiation tolerant up to 75~kRad Total Ionizing Dose (TID) and tested for SEU with a total fluence of \mbox{$3.8 \times 10^{12}$~$h$/cm$^2$}.
\item The energy measurement requires a dynamic range of  approximately 12 bits to digitize energies from 32~MeV to 102~GeV for the front layer trigger cells and from 125~MeV to 400~GeV in the middle layer trigger cells \cite{phase1tdr}.
\end{itemize}
This combination of speed, precision, low power, and particularly radiation hardness is not readily available commercially. The following sections describe a chip that meets these specifications. 

\subsection{Radiation Tolerance}
Though designed for Phase 1, the ADC is expected to continue to perform reliably throughout the operation of the HL-LHC. For the expected total integrated luminosity of 3000~fb$^{-1}$ the LAr electronics are required to be radiation tolerant to a TID of 75~kRad, and tested for Single Event Effects (SEE) with a total fluence of \mbox{$3.8 \times 10^{12}$~$h$/cm$^2$}~\cite{hllhcscoping}. The radiation at the LAr electronics comes predominantly from secondary particles produced by interactions of primary particles with the detector elements. The most significant contribution to the flux in the LAr electronics are photons and neutrons. Charged hadrons (mainly protons and pions) are a smaller contribution. 

A commercial 130~nm CMOS process was selected for this ADC. The rated supply voltage of thin (thick)-oxide 130~nm devices is 1.2~V (2.5~V). The 130~nm CMOS process has been demonstrated for both thin-oxide and thick-oxide MOS devices to be radiation hard to a few tens of MRads~\cite{rad_tol_ref1, rad_tol_ref2}.  It was noted~\cite{rad_tol_ref3} that thin-oxide devices are more radiation tolerant than thick-oxide devices. However, a previous design~\cite{nevis10} of a multiplying digital-to-analog converter (MDAC) using thick-oxide devices in this technology showed that standard hardness-by-design layout techniques were sufficient to achieve the necessary radiation tolerance. The thick-oxide devices allow for an input signal swing of the ADC of 2.4~V$_{p-p}$ differential, relaxing the noise requirements in the ADC. 

\subsection{Prototype Implementation}
\label{sec:arch}
\begin{figure}
	\begin{center}
		\includegraphics[scale=0.37]{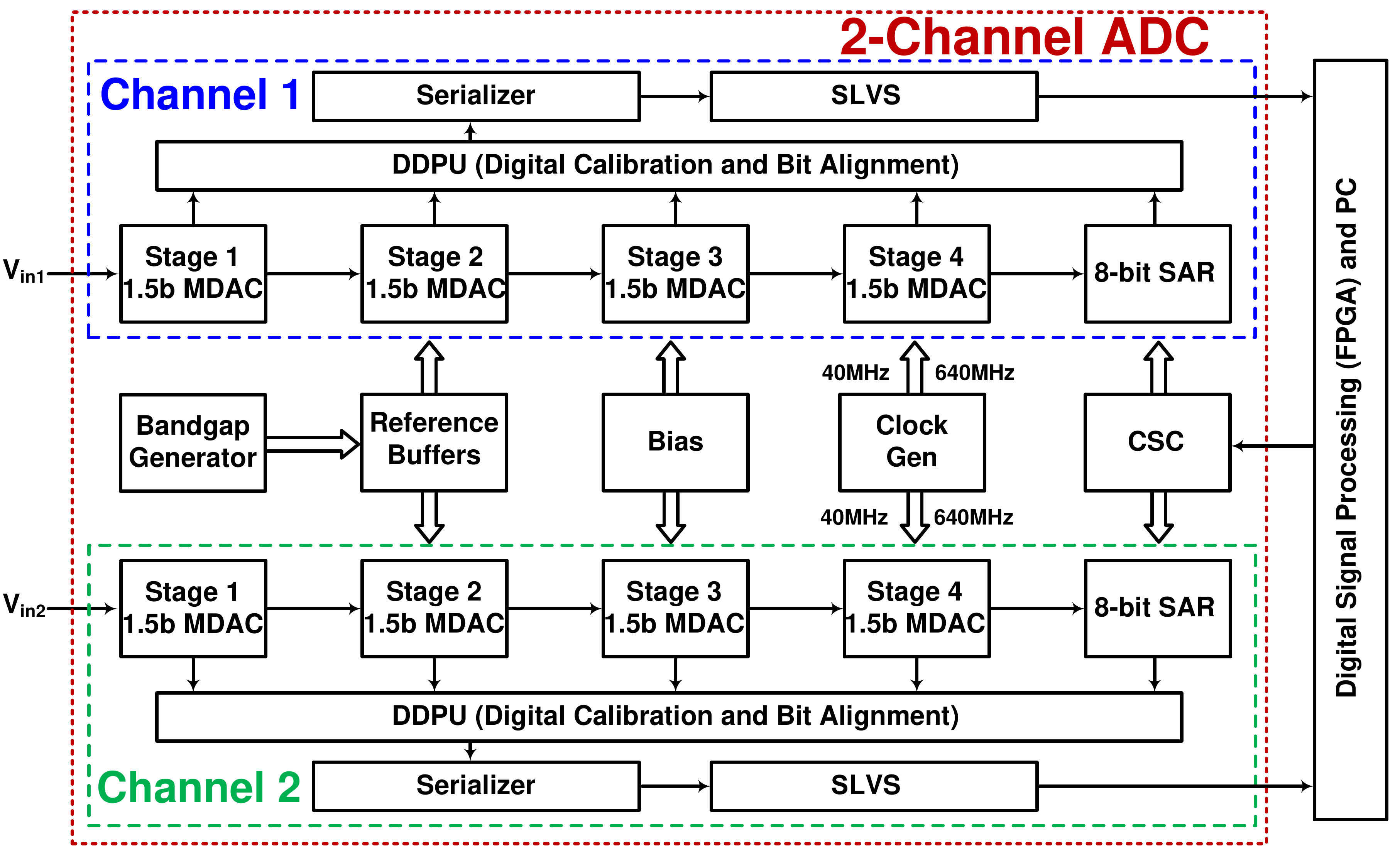}
		\caption{Prototype architecture.}
		\label{fig:chip_arch}
	\end{center}
\end{figure}

The goals of the current prototype are to test the performance of the complete ADC (extended from \cite{nevis10} with the addition of an 8-bit SAR backend), verify the on-chip, digital mismatch calibration and study the radiation tolerance of the ADC. The prototype's architecture, shown in Figure \ref{fig:chip_arch}, consists of two identical 12-bit 40 MS/s pipeline and SAR ADC channels. In 130~nm CMOS capacitor matching is not guaranteed to the precision required for a 12-bit ADC. The pipeline architecture allows for a calibration to correct gain errors due to the capacitor mismatch. The two ADC channels are fed with $V_{in1}$ and $V_{in2}$ respectively. 

Each of the two ADC channels in the current prototype consist of four MDAC stages with a gain of two each. Each MDAC stage resolves one bit, followed by a times two residue amplification. The three possible output codes (``1.5 bits'') allow for the digital error correction. The analog residue of the fourth and final MDAC stage needs to be further resolved to 8-bit accuracy to determine the final 12-bit ADC word. This analog residue is fed to the 8-bit SAR to determine the eight least significant bits. In the current prototype, the stage-1 MDAC is nominally 12-bit accurate, identical to the MDAC in~\cite{nevis10}. To reduce power consumption the following three MDACs are scaled. The scaling is achieved by reducing the silicon size and therefore the power used in these MDACs. The main building block of the MDACs is the Operational Transconductance Amplifier (OTA). The OTA for stages 2-4 requires less than the 12-bit precision of the first stage. Therefore, the open loop gain can be lower and the sampling capacitor size can be smaller, reducing the power consumed. The implementation details of the ADC are further explained in the following sections.

\subsubsection{MDAC and OTA Implementation}

\begin{figure}
	\begin{center}
		\includegraphics[scale=0.5]{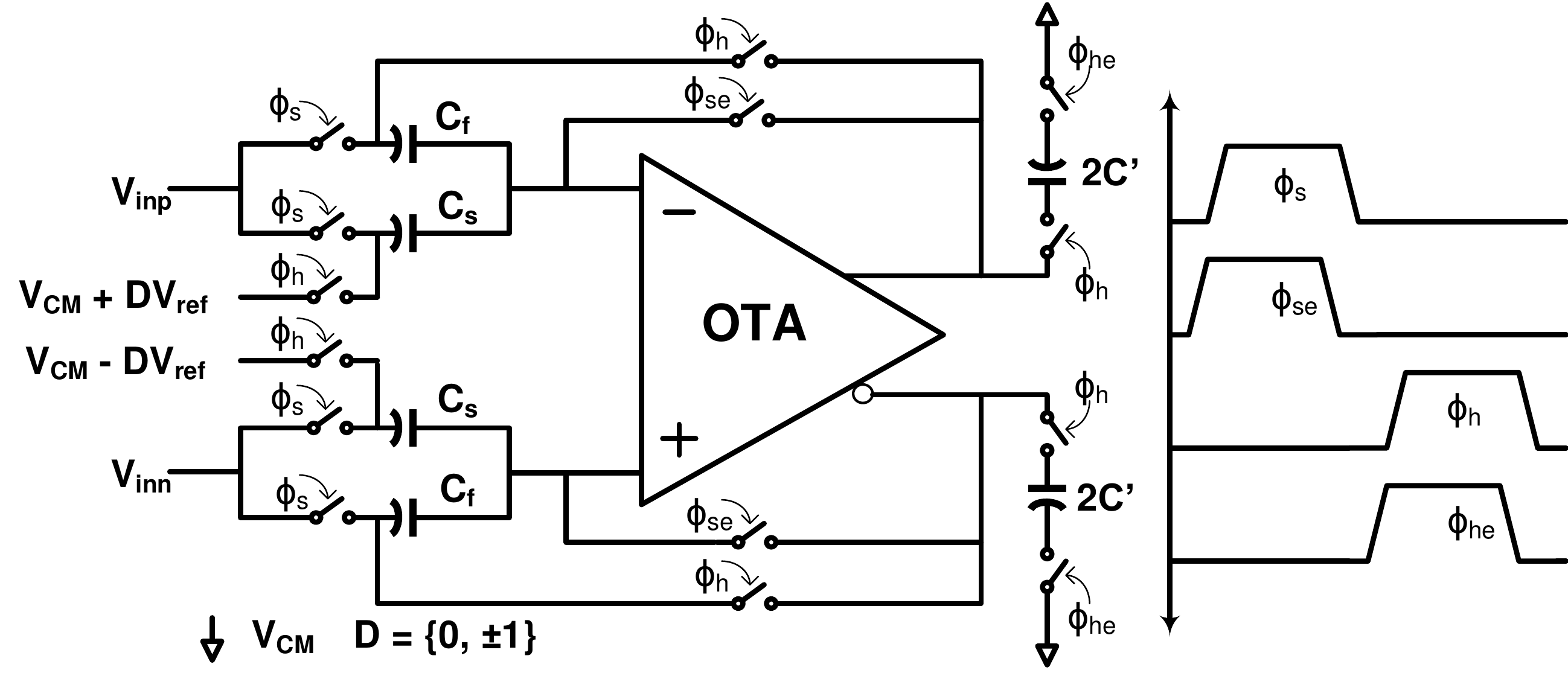}
		\caption{1.5b MDAC stage (subADC not shown); \mbox{$\phi_s/\phi_h$ - sample/hold} phase; \mbox{$\phi_{se}/\phi_{he}$ - advanced} version of $\phi_s/\phi_h$; \mbox{$V_{CM}$ - common-mode} voltage; \mbox{$V_{ref}$ - reference} voltage; \mbox{$D$ - subADC} decision.} 
		\label{fig:MDAC}
	\end{center}
\end{figure}

Figure \ref{fig:MDAC} shows the architecture of the 1.5-bit MDAC stage and OTA used in the pipeline ADC \cite{stg_arch}. This architecture was described in detail in a previous paper \cite{nevis10}.  The MDAC consists of a flip-around architecture with the input $V_{in}$ sampled onto the sampling capacitors $C_s$ and the flip-around capacitors $C_f$, during the sampling phase $\phi_s$. Nominally, $C_s$ and $C_f$ are equal.  At the beginning of the hold phase $\phi_h$, depending on the subADC decision $D$, the reference $V_{ref}$ is appropriately connected to the capacitors, thus performing the required MDAC operation given by

\begin{equation}
        V_{out} = 2V_{in} - DV_{ref}
        \label{eqn:MDAC_opn}
\end{equation}
where $D = {0,\pm1}$ is the subADC decision. The subADC (not shown) consists of a flash comparator and an input sampling network. The OTA consists of a single stage, gain-boosted, folded cascode amplifier with an NMOS input pair. The OTA is designed for a DC gain of 80~dB (first stage) and a UGB of 450~MHz, thus providing enough margin for the targeted 12-bit performance at 40 MS/s.

\subsubsection{SAR Implementation} \label{sec:SAR}
\begin{figure}
	\begin{center}
		\includegraphics[scale=0.5]{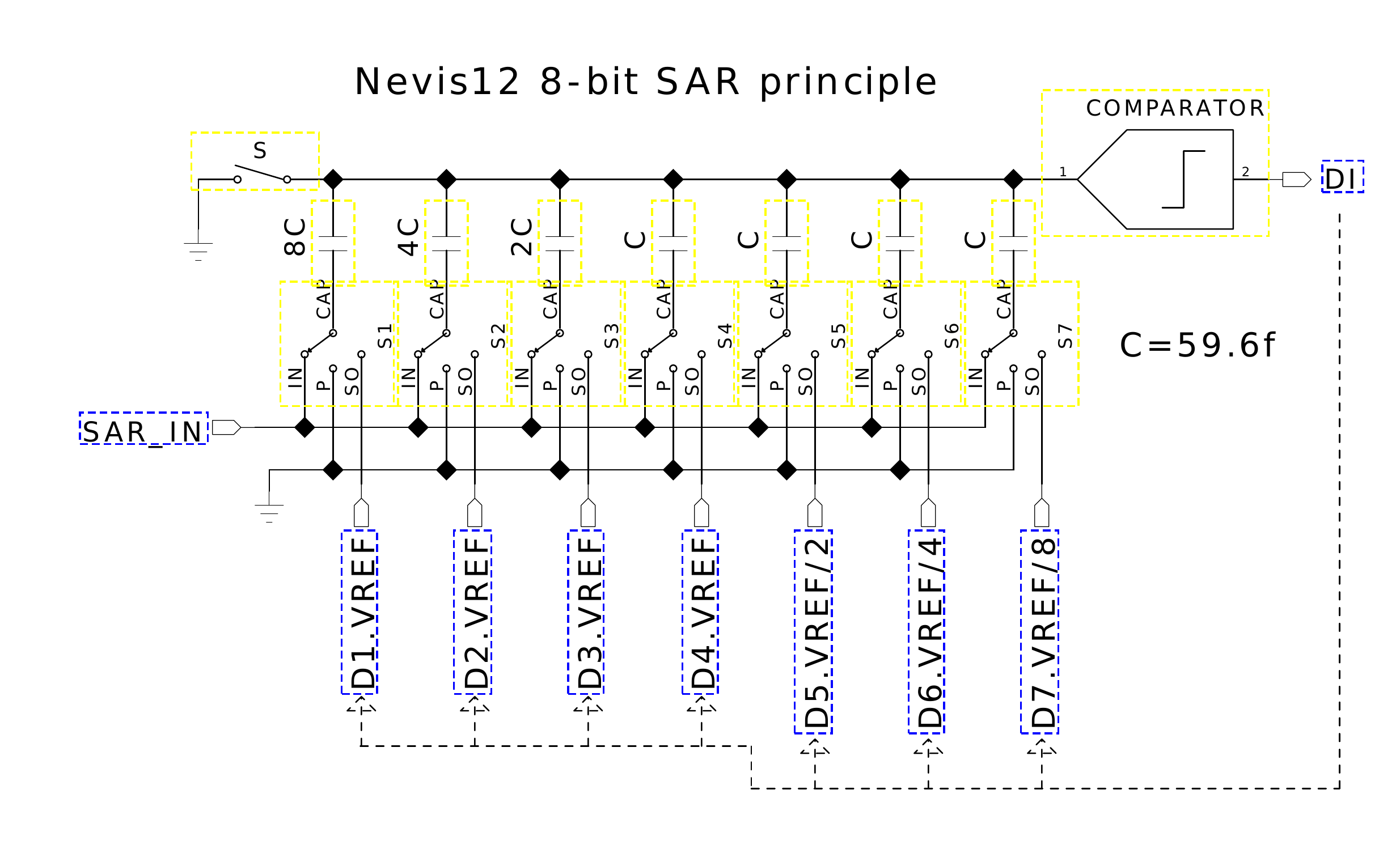}
		\caption{The SAR architecture block diagram. A single-ended version is shown for clarity, whereas the actual implementation is fully differential.}
		\label{fig:sar_arch}
	\end{center}
\end{figure}

The eight least significant bits of the ADC are resolved using a SAR. A synchronous SAR design was chosen for simplicity and the relatively moderate speed and resolution requirements (8-bit and 40~MS/s). Figure~\ref{fig:sar_arch} shows the simplified block diagram of the 8-bit SAR, with the digital logic not shown. A single-ended version is shown for simplicity, whereas the actual implementation is fully differential. 

The selection of the total SAR input capacitance is a compromise between the radiation tolerance, noise and mismatch. The required radiation tolerance necessitates the use of larger unit capacitor sizes than dictated by the 8-bit noise and matching requirements. The SAR uses a unit capacitor size of 59.6~fF. This capacitor size is smallest MiM (Metal-insulator-Metal) capacitor with best mismatch in the technology used. To reduce the total capacitance of the SAR with the chosen unit capacitor size, the SAR capacitor array is segmented, as shown in Figure~\ref{fig:sar_arch}. The first four largest capacitors are binary scaled with a total capacitance of $15C$, while the final three capacitors have the same unit capacitance $C$. To enable binary successive approximation, the final three capacitors are fed with the scaled reference voltages of ($\pm V_{ref}/2, \pm V_{ref}/4, \pm V_{ref}/8$) obtained through a resistive ladder. 

During the sampling phase, the residue from the stage four MDAC is sampled into the capacitor array. The switches $S1$ to $S7$  are in the ``IN'' position (see Figure~\ref{fig:sar_arch}). The SAR conversion process starts with turning switch $S$ off and switches $S1$ to $S7$ to the position ``P,''  when all the capacitors are pulled to $V_{CM}$ (0.6~V) and the most significant bit ($D_1$) is resolved. Sequentially, as a result of the comparison, switches $S1$ to $S7$ are moved to position ``SO'' based on the decisions $D_i$ to perform the successive approximation operation by adding or subtracting charge from input signal. 

\begin{figure}
	\begin{center}
		\includegraphics[scale=0.4]{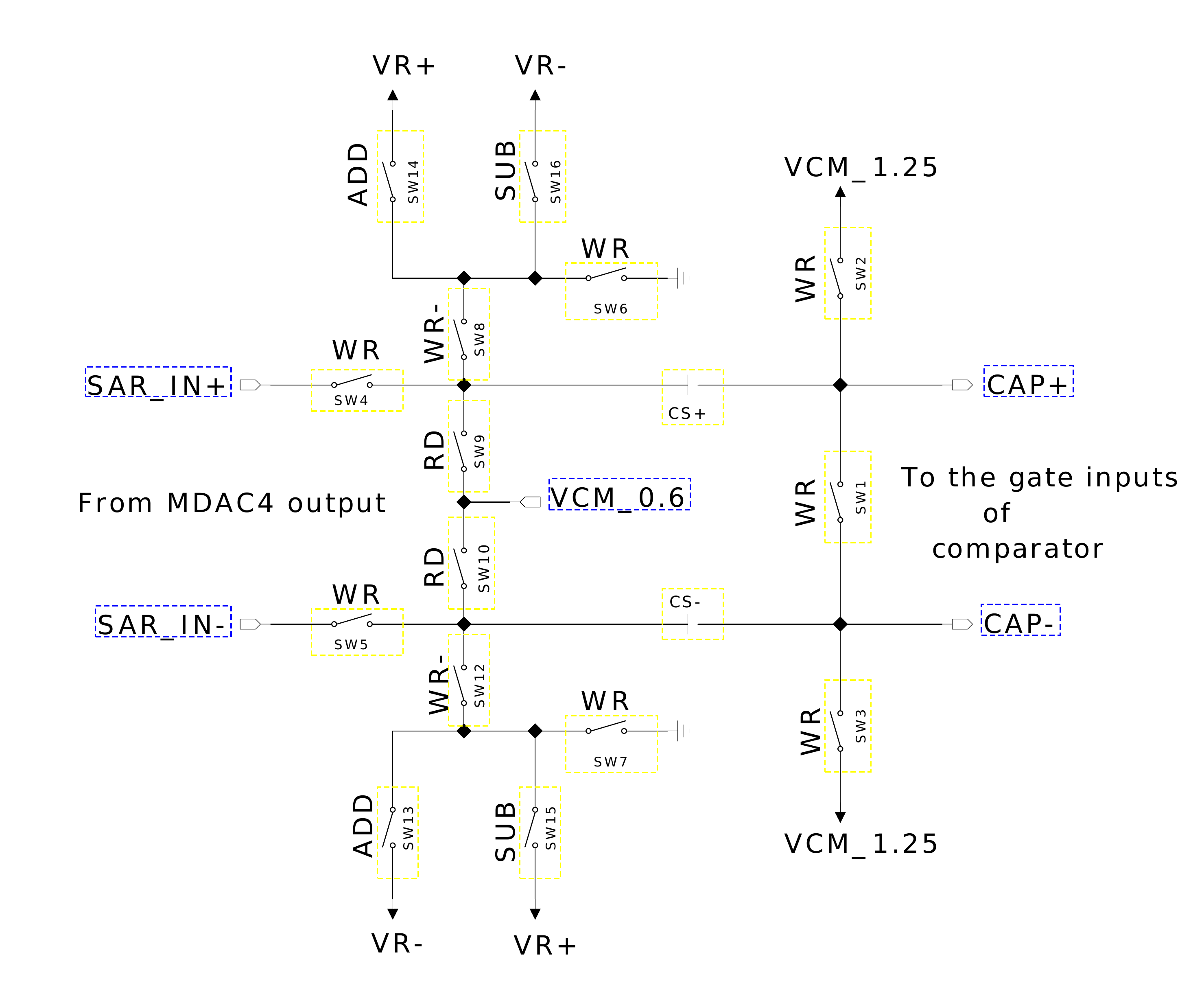}
		\caption{Detail of switches for the voltage shift network, necessary at the MDAC-SAR interface when the SAR (1.2~V supply) samples the input MDAC signal (2.5~V supply).}
		\label{fig:sar_switch}
	\end{center}
\end{figure}

The MDAC stages, implemented using thick-oxide MOS devices, operate from a supply voltage of 2.5~V and a common-mode of 1.25~V. The 8-bit SAR is implemented using thin-oxide MOS devices in a supply voltage of 1.2~V with a common-mode of 0.6~V for low-power operation (SAR power consumption is $\approx fCV_{DD}^2$ where $V_{DD} = 1.2$~V). This realizes the SAR digital logic using available standard thin-oxide cells. As a result of the different voltage domains of the MDACs and the SAR, a level shift network is necessary at the MDAC-SAR interface when the SAR samples the input signal.  Figure~\ref{fig:sar_switch} shows the detailed implementation of the DC level-shift network incorporated in the switches $S1$ to $S7$ that accomplishes this operation.  

The SAR input sampling phase is simultaneous with the hold phase of MDAC4. While the $WR$ signal is active, switches $SW4$ and $SW5$ (in Figure~\ref{fig:sar_switch}) sample the input signal onto one plate of the capacitors. Meanwhile, switches $SW1-3$ short the other plate to 1.25~V, the common-mode voltage of the MDACs. During the sampling time the thin-oxide switches $SW13-16$ are protected from any over-voltage by switches $SW8$ and $SW12$. After the sampling phase the left plate of the capacitors is pulled down to 0.6~V using switches $SW9 -10$ (while the $RD$ signal is active). This accomplishes the common-mode translation from 1.25~V to 0.6~V. At the same time, the comparator compares the voltage on its $CAP$ inputs and produces the SAR digital output for a given bit. The result is applied in the next 640~MHz clock tick by adding or subtracting the reference voltage from the sampled signal. The SAR unit is operated in pipeline mode with the MDACs. The complete SAR operation requires 12.5~ns (one half of one sampling period). The SAR digital control of switches and outputs is implemented using standard 1.2~V logic, with standard D flip-flops (the primary state machine memory elements) clocked at 640~MHz.

\subsubsection{Digital Data Processing Unit and Digital Gain Calibration} \label{sec:ddpu}
\begin{figure}
	\begin{center}
		\includegraphics[scale=0.37]{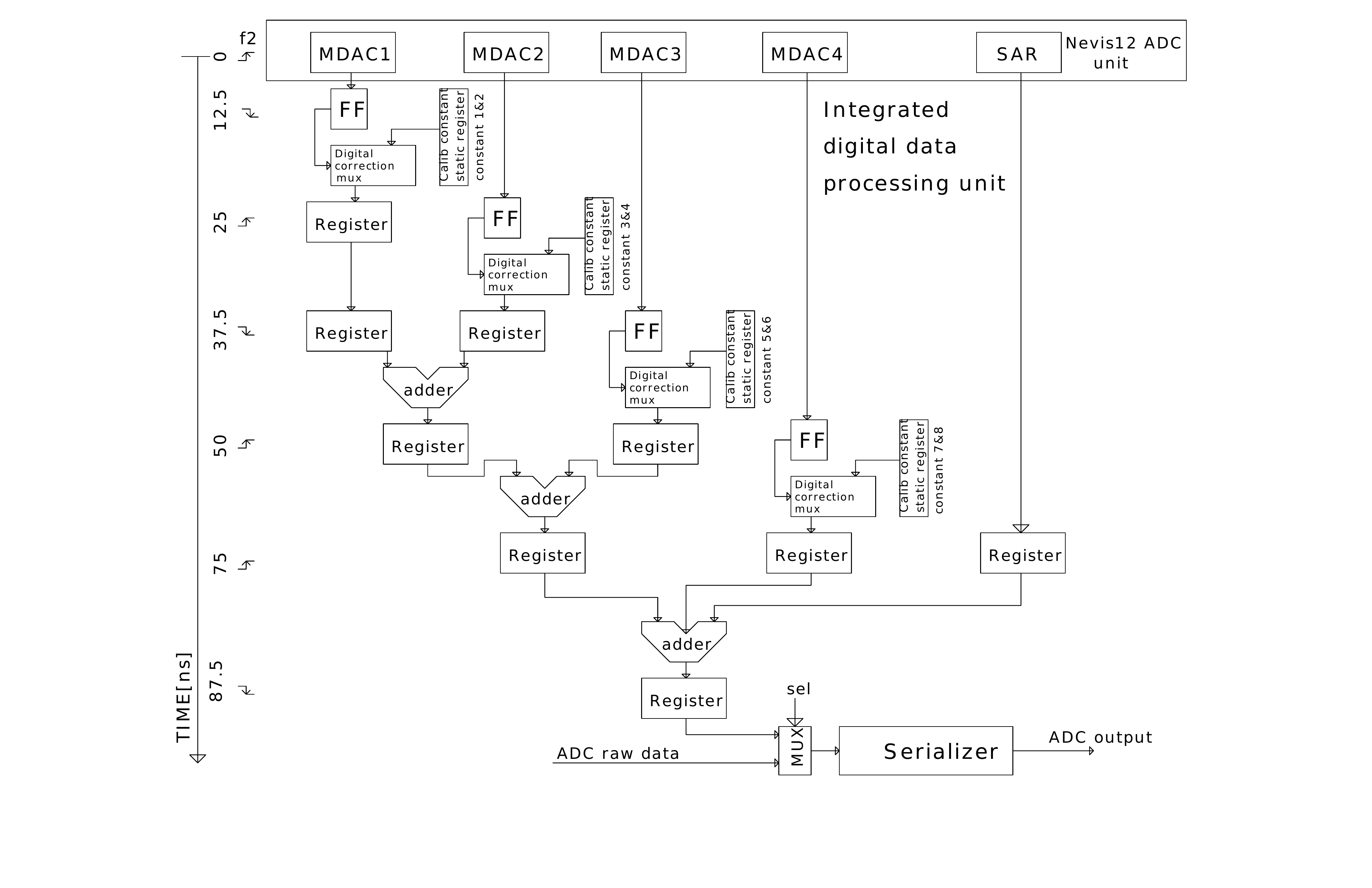}
		\caption{DDPU block diagram. The DDPU applies the calibration to the digital output of each MDAC and forms the final ADC word.}
		\label{fig:ddpu}
	\end{center}
\end{figure}

Digital signals from the MDACs and the SAR are processed by the digital data processing unit (DDPU). The DDPU aligns the data in time, performs the digital correction and forms the A/D result. In Figure~\ref{fig:ddpu} the time structure of the digitization is shown, with a latency (to first bit read out) of 87.5~ns. The digital output of each ADC pipeline stage is produced sequentially using both edges of the sampling clock. In order to form the final ADC word such an output must be aligned: this is done by registering the data in the DDPU and adding the correct number of registers before summing.

\begin{figure}
	\begin{center}
		\includegraphics[scale=0.8]{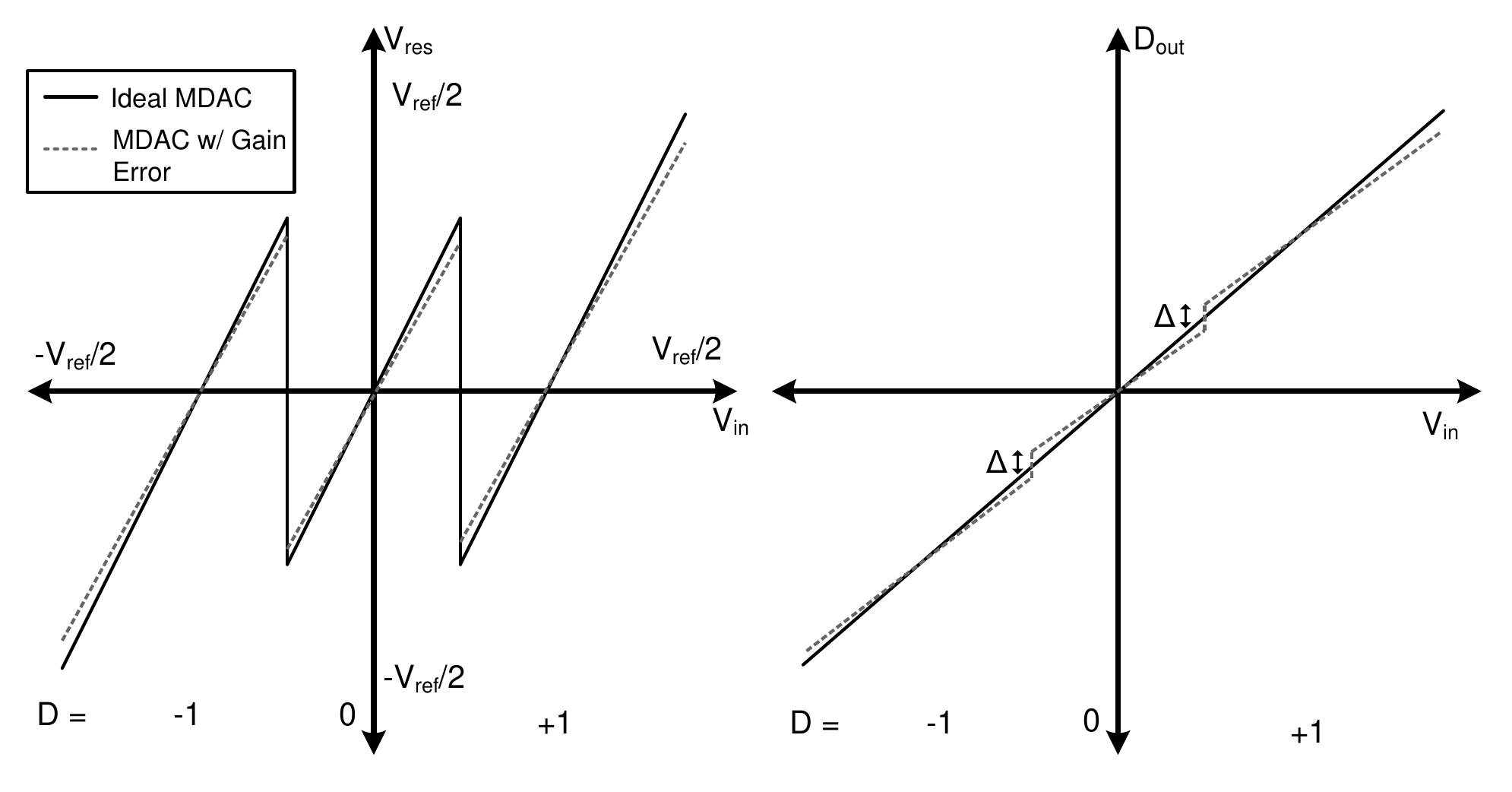}
		\caption{MDAC residue characteristic (left). Reconstructed output (right); $V_{ref}$ - reference voltage; $V_{res}$ - residue voltage; $D$ - subADC decision; $D_{out}$ - reconstructed output.}
		\label{fig:residue_char}
	\end{center}
\end{figure}

The most important function of the DDPU is the MDAC digital gain correction. Foreground digital calibration is performed to correct for gain errors due to capacitor mismatch~\cite{calib_ref}, exploiting the redundancy offered by the three possible codes produced to resolve one bit. Figure \ref{fig:residue_char} shows the residue output $V_{res}$ of the stage 1 MDAC and the reconstructed output $D_{out}$, as a function of the input $V_{in}$,  for an ideal MDAC and for an MDAC with gain error. Gain errors due to capacitor mismatch give rise to code jumps in the reconstructed output, as shown by $\Delta$ in Figure \ref{fig:residue_char}. The calibration algorithm consists of measuring the MDAC code jumps $\Delta$ by subsequent ADC stages and removing them digitally from the reconstructed digital output $D_{out}$. The calibration procedure starts with the last MDAC stage (the stage 4 MDAC in this prototype) and moves backward to calibrate the stage 1 MDAC. An on-chip control register is used to put the ADC in calibration mode. Once determined, the eight 12-bit calibration constants are stored in a triple-redundant buffer by the chip calibration and control system. 

The final ADC word is formed with the 3-way adder when the corresponding SAR output is available. This 12-bit word is sent off-chip in serial form at 640~MHz. In total, 16-bits of serial data are sent, usually with four bits set to zero. The DDPU can also send out the uncorrected digital data from each MDAC and SAR, using all 16-bits.  The DDPU unit is implemented using fully synthesized logic using standard CERN scripts for commercial 130~nm digital libraries with Verilog input. The DDPU operates at 80~MHz and consumes~$\sim$~3~mW of power at 1.2~V. 

\subsubsection{Clock Generator Implementation} \label{sec:clk}

\begin{figure}
	\begin{center}
		\includegraphics[scale=0.4]{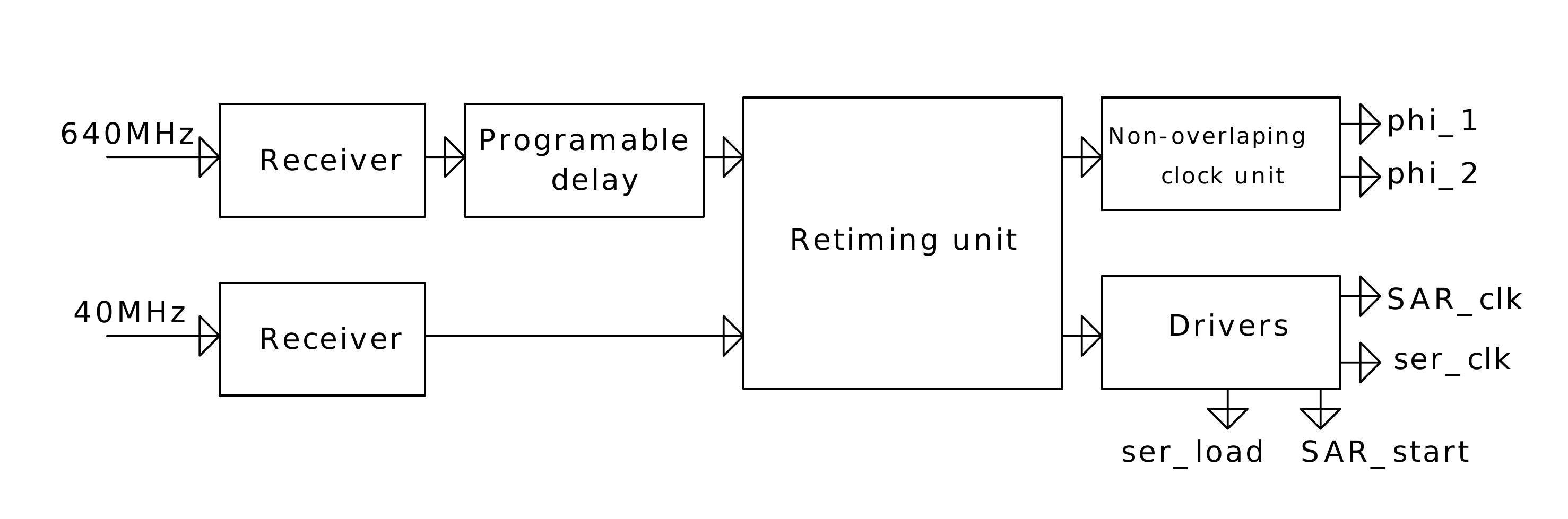}
		\caption{Clock architecture block diagram.}
		\label{fig:clk_arch}
	\end{center}
\end{figure}

A 640~MHz sinusoidal clock is fed into the chip and divided down to generate the required 40~MHz clock phases, while a separate 40~MHz clock provides the frame. The clock signal and time-critical control signals are produced inside the clock generator unit (Figure~\ref{fig:clk_arch}). The reliable transmission of serialized output requires either a precise and stable clock or a Phase Lock Loop (PLL).  A precise clock is also critical at the interface between MDAC stages (operating at 40 ~MHz) and the SAR stage (operating at 640~MHz). As an alternative to operating an on-chip PLL in a radiation environment, this ADC tests the on-board distribution of 40~MHz and 640~MHz clocks. Both clock signals are distributed using the scalable low-voltage signaling (SLVS) protocol~\cite{slvs_spec}. The timing information is expected to be on the positive edge of the 640~MHz signal; therefore, the receiver of this signal must meet the jitter requirements for a 12-bit ADC ($\sim$~1~ps). This implementation does not expect the clock signals to be synchronized, and provides the ability to adjust the clock relationship using a programmable delay (up to $\sim$~3~ns for the 640~MHz clock).

The 40~MHz clock provides the frame for the sampling. The edge of the frame is detected by the 640~MHz clock, and the information passed to a 16-bit shift register. This arrangement allows the straightforward generation of the ADC sampling phases $\phi_1$ and $\phi_2$ with a simple RS flip-flop. The ability to adjust the start of the $\phi_1$ and $\phi_2$ signals by up to eight 640~MHz clock steps using a programmable delay is provided. The same shift register also distributes the timing critical signals for loading the serializer and starting the SAR operation, and this location contains the fan-out of the fast clocks inside the ADC (640~MHz for SAR operation and 640 MHz for serialization). 

\subsubsection{Calibration and System Control } \label{sec:csc}
The calibration and system control block (CSC) provides the slow control for all systems on the chip. The CSC uses a three wire serial interface (5~MHz clock, data$_{in}$ and frame$_{in}$) to write to the chip registers and a two wire serial interface (data$_{out}$, frame$_{out}$) to read them back. The CSC communicates with the DDPU units inside the chip using the same serial protocol. The control signals are retimed internally and the destination address within the DDPU is added to the external protocol.  Output data from the DDPU are multiplexed to two wires, which allows the loading of all calibration parameters and internal MDAC signals needed to perform the calibration. The CSC unit contains the triple redundant memory (to protect against SEU) for the chip configuration and is a fully synthesized block using the 1.2~V standard library.

\subsubsection{Serializer} \label{sec:serializer}
\begin{figure}
	\begin{center}
		\includegraphics[scale=0.40]{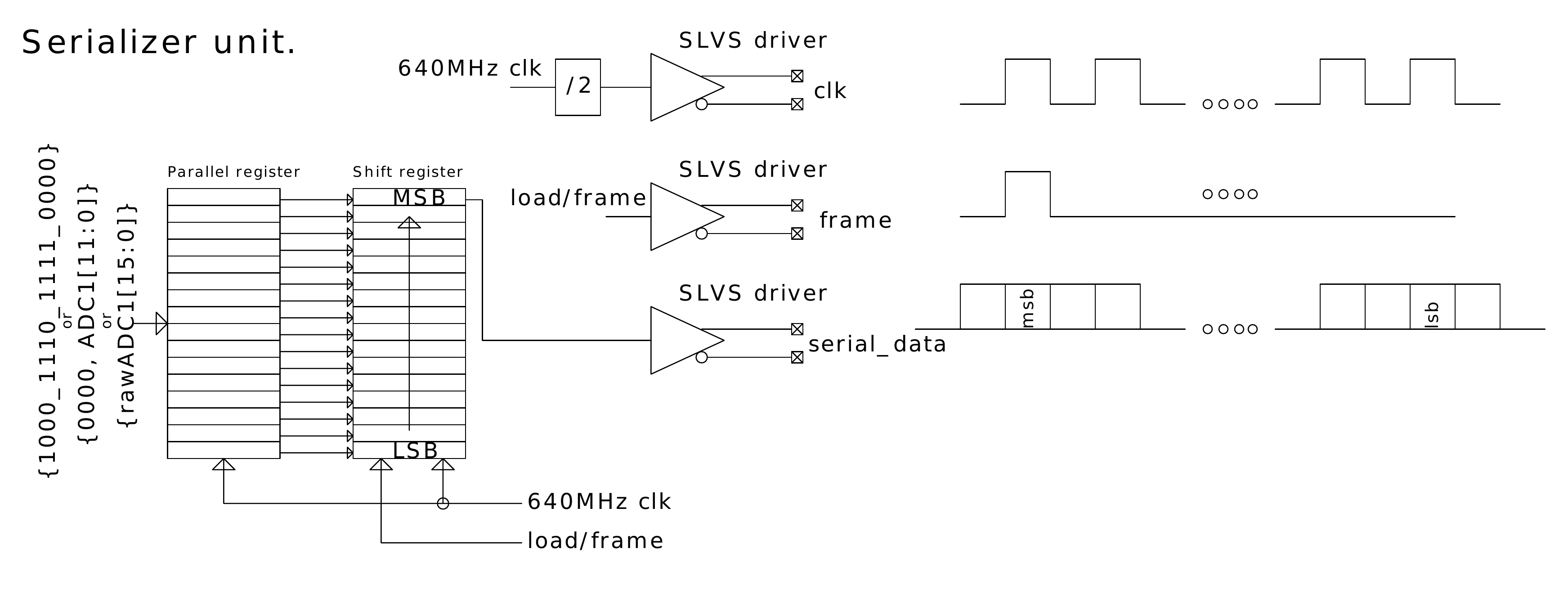}
		\caption{Serializer block diagram showing the parallel and shift registers.}
		\label{fig:serializer}
	\end{center}
\end{figure}

The serializer unit (Figure~\ref{fig:serializer}) consists of a parallel register and a shift register. With each 640~MHz clock signal, the data are written into a parallel register.  In normal operation the unit can send out calibrated ADC data, or alternatively a test pattern or uncorrected data.  At the end of the shift operation the parallel register data are loaded into the serial register and shifted outside the chip. The start of the new shift operation is marked by the frame signal, which is also sent with the data. A 320~MHz clock signal is sent in parallel with the serial ADC data. The clock signal is produced locally in the serializer by dividing the 640~MHz clock. Each edge of the 320~MHz clock signal is used to strobe the data in the off-chip de-serializer unit. The serializer unit is implemented using the standard 1.2~V library. The ADC data, clock signal and the frame are driven outside the chip using SLVS drivers designed at CERN~\cite{slvs_cern}. To ensure radiation hardness triple redundant logic is used in the following manner: If the D-flip flops are part of the state machine, or if they store the calibration or other slow control bits, then they are tripled and the result is generated by voting. D-flip flops where the outputs are recomputed from scratch by up to two free running clock cycles are not triple redundant.

\section{Measurement Results} \label{sec:meas_results}

\begin{figure}
	\begin{center}
		\includegraphics[scale=0.25]{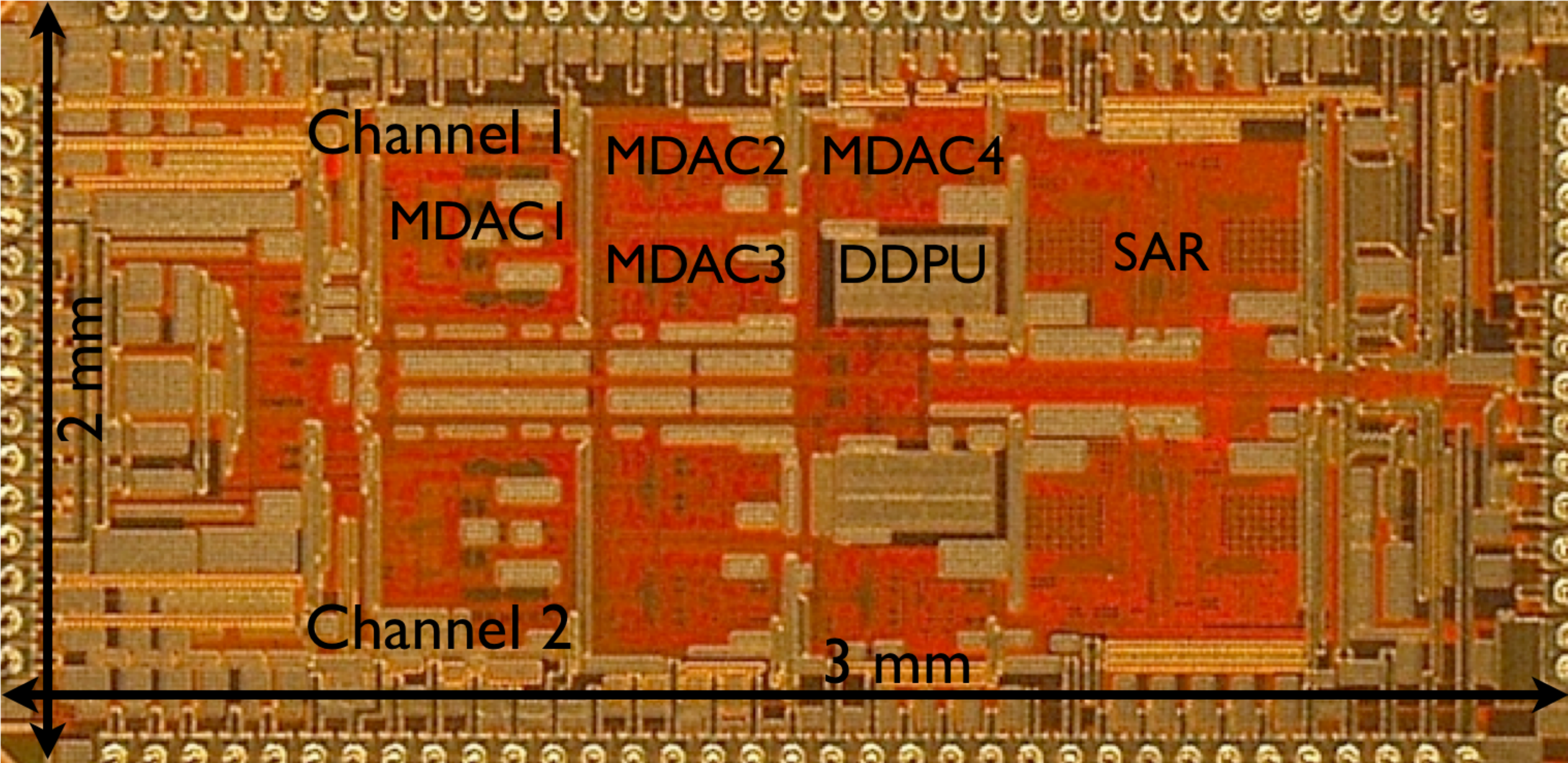}
		\caption{ADC die photograph.}
		\label{fig:die_photo}
	\end{center}
\end{figure}

Figure \ref{fig:die_photo} shows a die photograph of the 2-channel ADC prototype. The chip occupies 6~mm$^{2}$. To characterize the dynamic performance of the ADC, an input sinusoidal signal is filtered to remove harmonic components and fed to the ADC input. To allow for flexibility in testing, the signal can be sent either to the ADC input, or bypassing the MDACs directly to the SAR ADC input. This feature allows the testing of the SAR stage as a separate 8-bit ADC. The digitized output can either be sent directly off-chip, or the calibration can be applied. The digital outputs from the prototype chip are collected by an FPGA. Thirty chips were packaged and functional tests were performed with a socketed test board, with a yield of approximately 90\%. The socketed board limited the precision of the measurements. Therefore, to determine ADC performance, two of the functional chips were soldered onto a precision test board. 

\subsection{ADC Performance} \label{sec:pre_rad}
\begin{figure}
\centering
\begin{minipage}{.45\linewidth}
	\includegraphics[width=\linewidth]{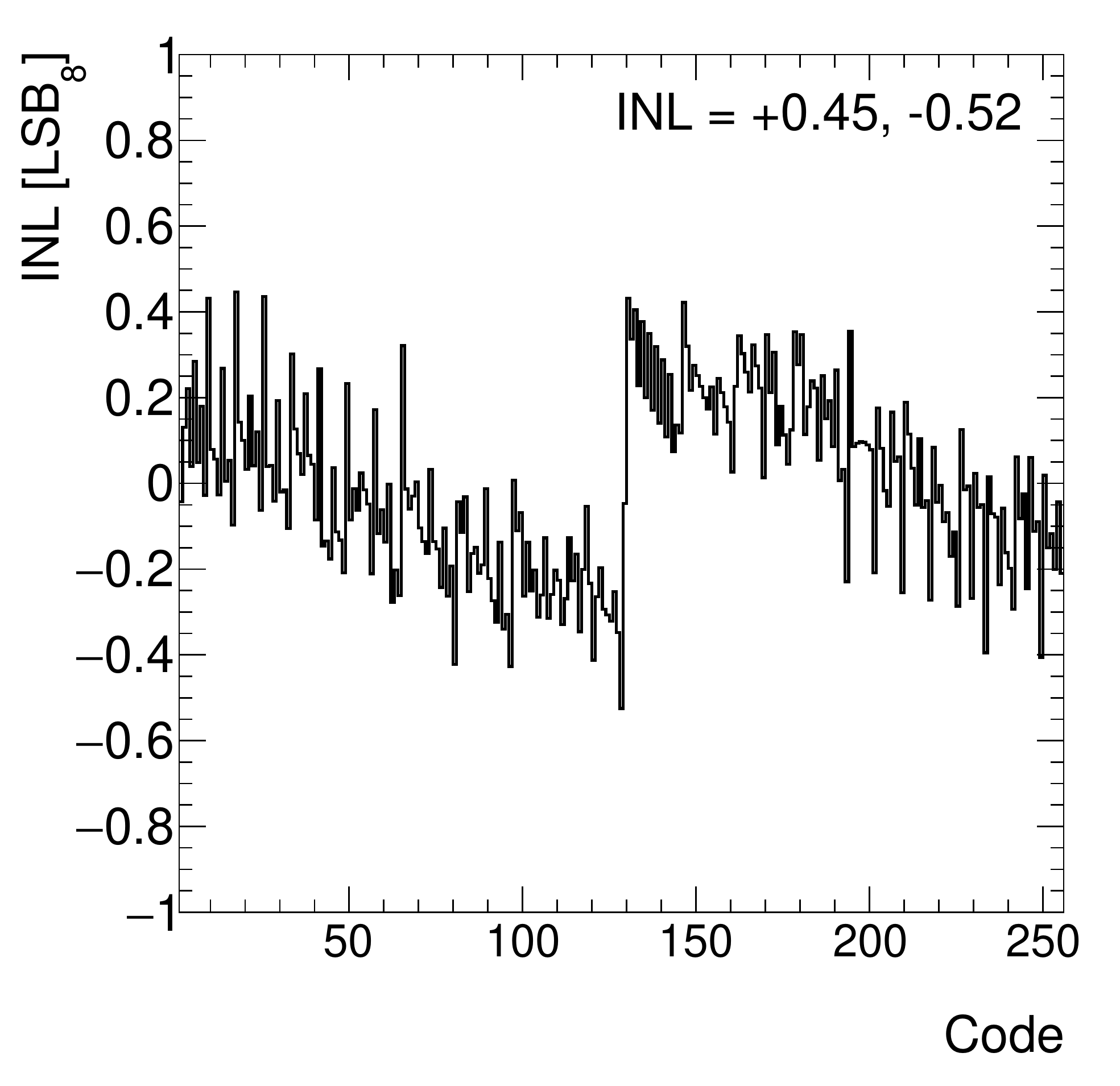}
	\caption{8-bit SAR sub-ADC INL at 40 MS/s ({\it Chip 1}).}
	\label{fig:8inl_full}
\end{minipage}
\hspace{.05\linewidth}
\begin{minipage}{.45\linewidth}
	\includegraphics[width=\linewidth]{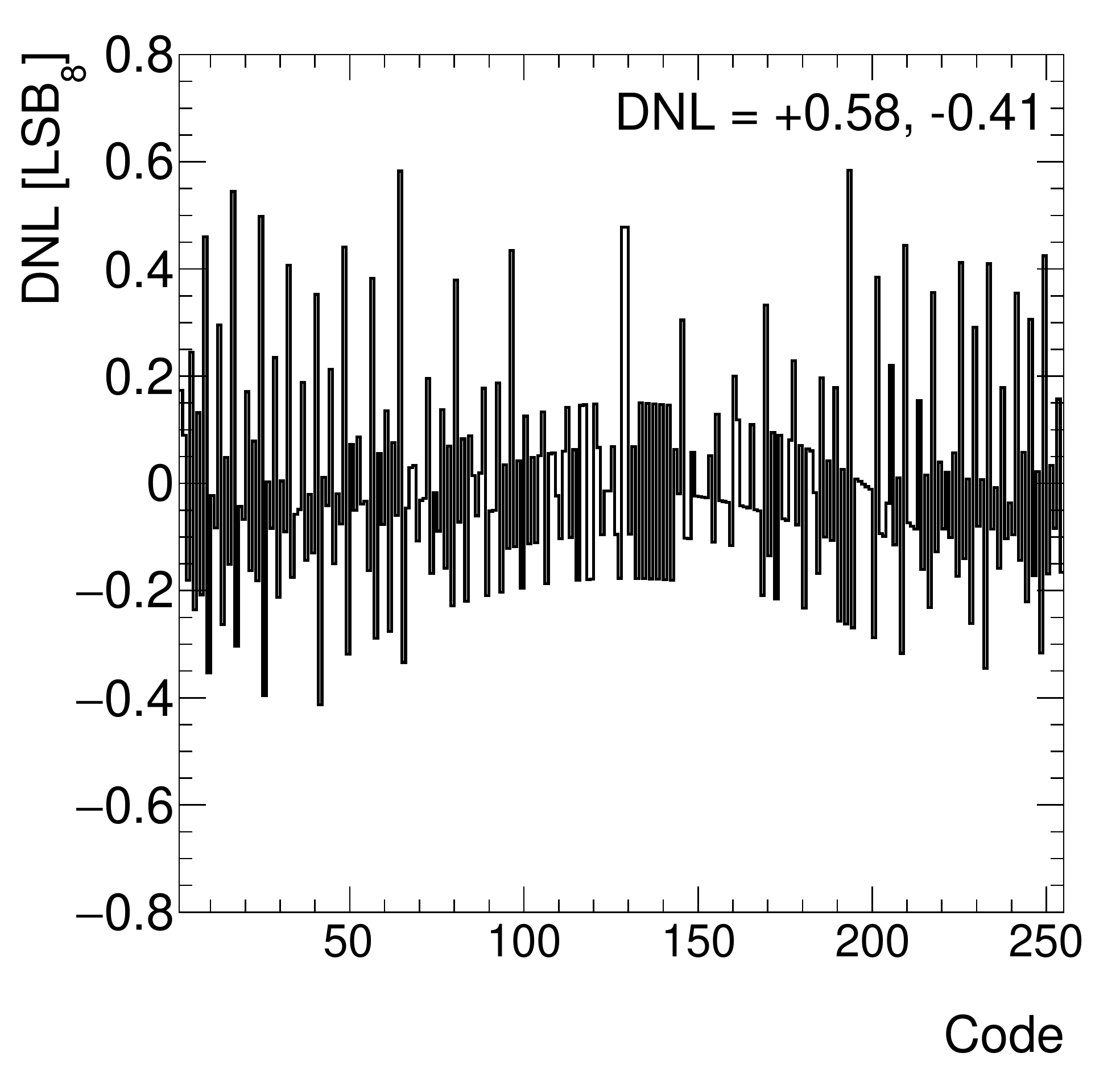}
	\caption{8-bit SAR sub-ADC DNL at 40 MS/s ({\it Chip 1}).}
	\label{fig:8dnl_full}
\end{minipage}
\begin{minipage}{.44\linewidth}
	\includegraphics[width=\linewidth]{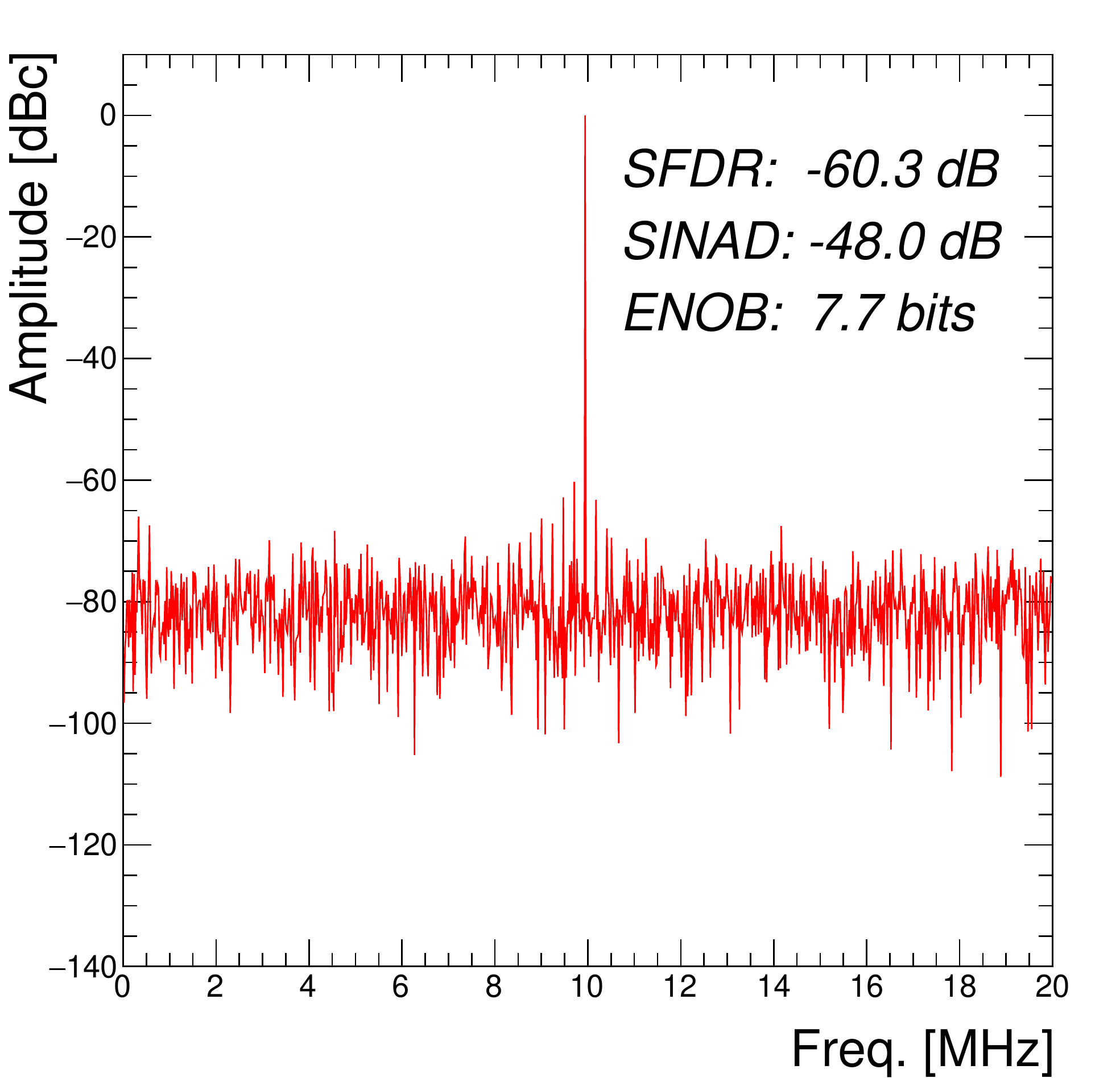}
	\caption{8-bit SAR sub-ADC FFT for \mbox{$f_{in}$ = 10~MHz} ({\it Chip 1}).}
  	\label{fig:8full_fft}
\end{minipage}
\end{figure}

Both ADC channels of the two chips mounted on the precision test boards were characterized. The 8-bit SAR ADC was first characterized independently of the MDACs. This was a critical test of the SAR and a major goal in designing this chip. The static performance of the ADC was measured by the sine-wave histogram method \cite{sine_inl_ref}. Figures \ref{fig:8inl_full} and  \ref{fig:8dnl_full}  show the typical static performance of one 8-bit ADC channel at 40~MS/s. The integral nonlinearity (INL) is +0.45/-0.52 LSB$_{8}$, where LSB$_{8}$ refers to the Least Significant Bit (LSB) of 8-bits. The differential nonlinearity (DNL) is +0.58/-0.41 LSB$_{8}$. Figure \ref{fig:8full_fft} shows the output fast Fourier transform (FFT) of a 10~MHz input sine-wave. The effective number of bits (ENOB) at 10~MHz is 7.7-bits and the spurious-free dynamic range (SFDR) is 60.3~dB. This performance was typical for all chips that passed the functionally tests, as the socketed board did not limit the precision of these tests. 

\begin{figure}
\centering
	\begin{minipage}{.45\linewidth}
		\includegraphics[width=\linewidth]{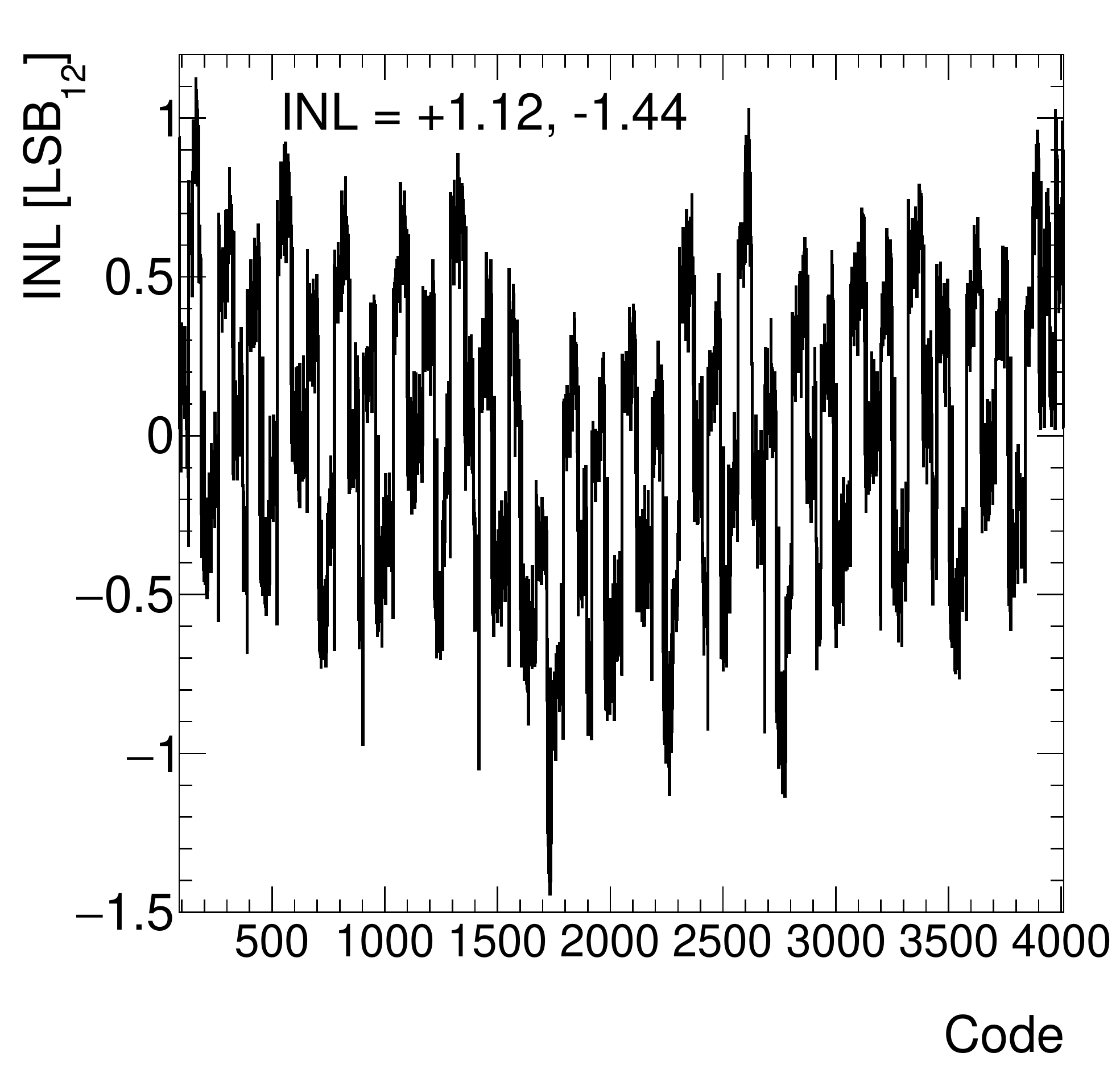}
		\caption{Full ADC INL at 40 MS/s ({\it Chip 1}): after calibration.}
		\label{fig:inl_full}
	\end{minipage}
	\hspace{.05\linewidth}
	\begin{minipage}{.45\linewidth}
		\includegraphics[width=\linewidth]{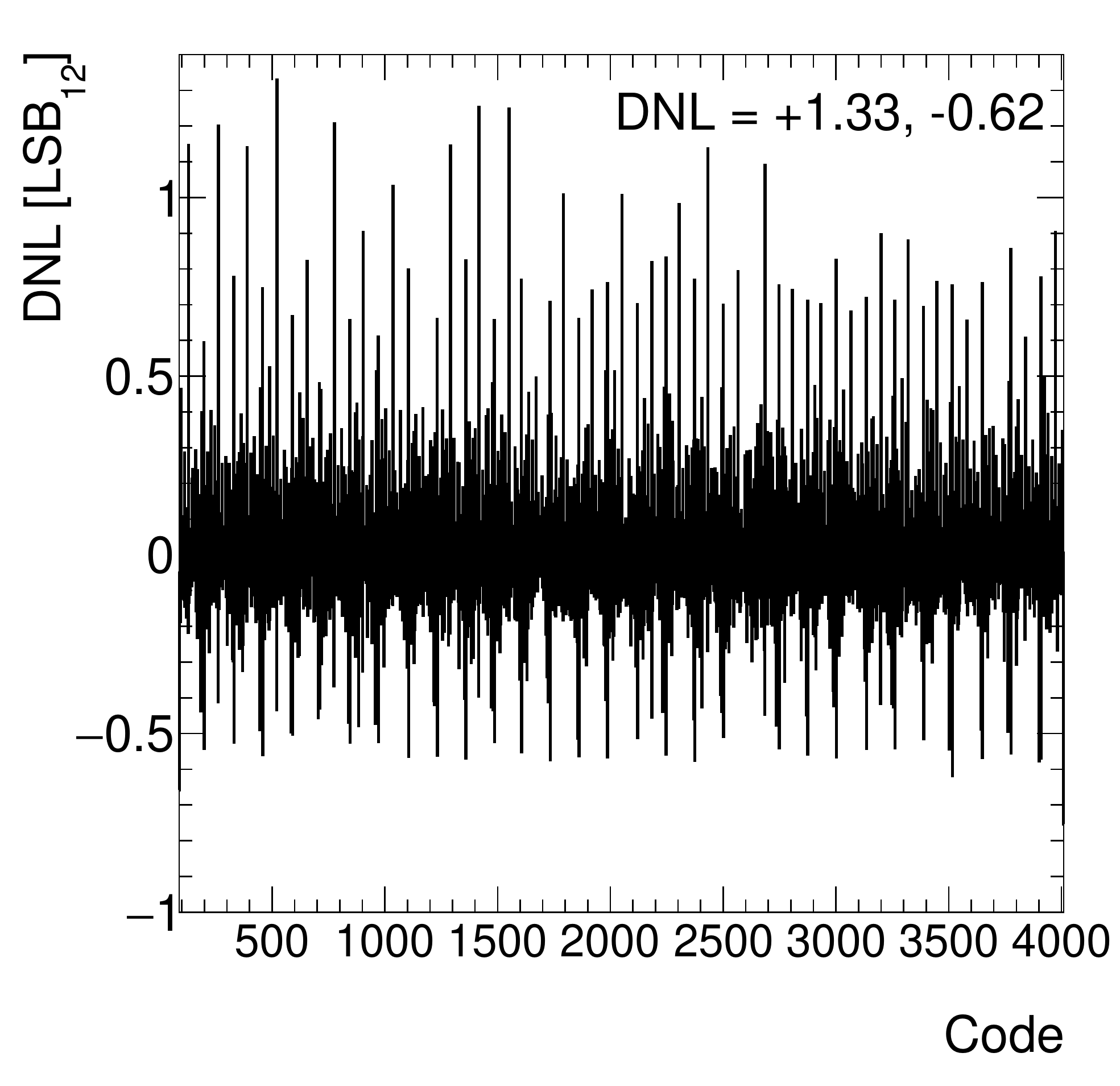}
		\caption{Full ADC DNL at 40 MS/s ({\it Chip 1}): after calibration.}
		\label{fig:dnl_full}
	\end{minipage}

	\begin{minipage}{.4\linewidth}
		\includegraphics[width=\linewidth]{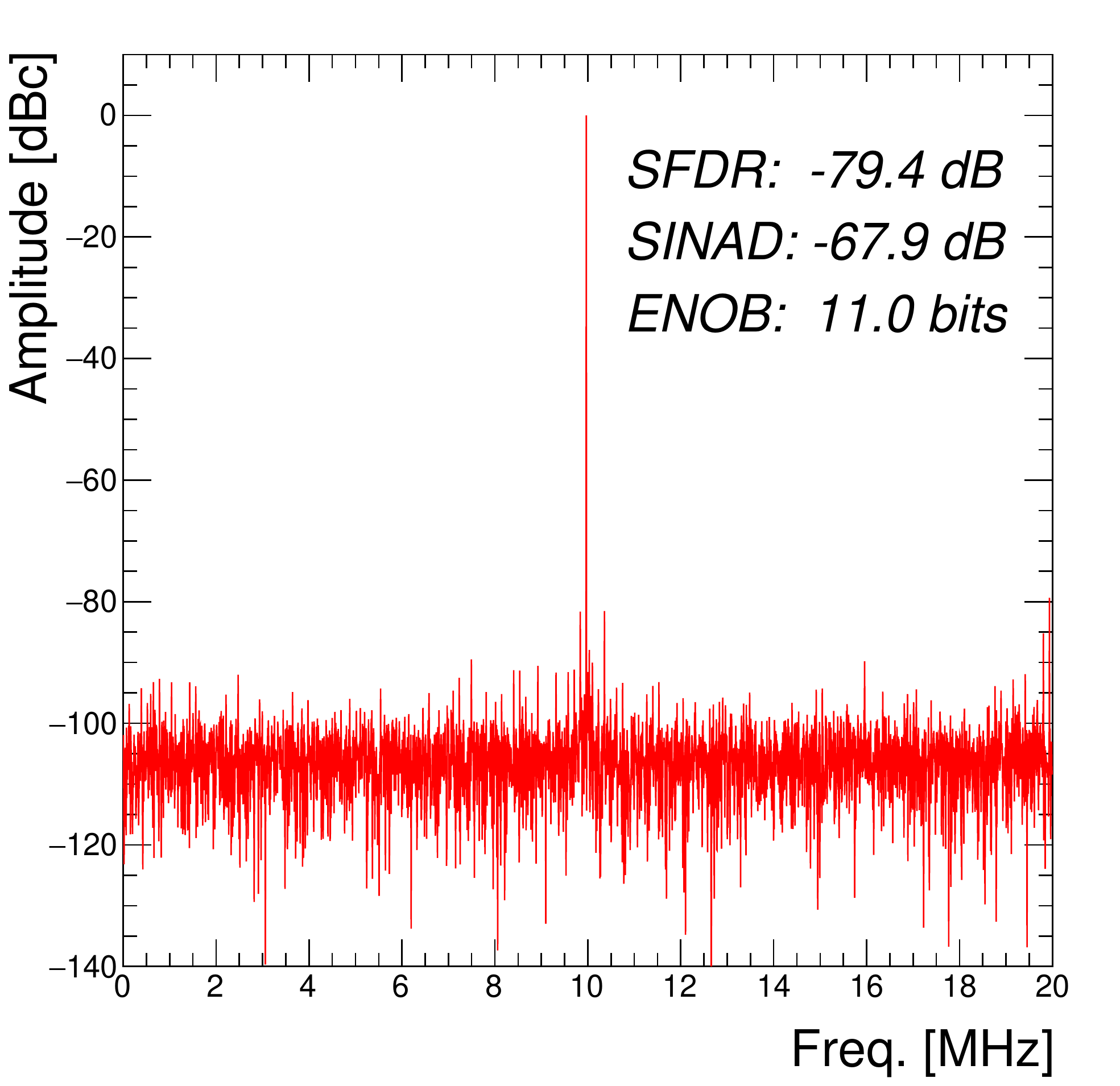}
		\caption{Full ADC FFT for \mbox{$f_{in}=10$~MHz} ({\it Chip 1}): after calibration.}
		\label{fig:full_fft}
	\end{minipage}
\end{figure}

The SAR stage is used to calibrate the MDAC stages, and the performance of the full 12-bit ADC is measured. Figures \ref{fig:inl_full} and  \ref{fig:dnl_full}  show the static performance of one 12-bit ADC channel at 40~MS/s. Similar performance is observed on both channels for both chips and crosstalk between channels is not observed. The INL is  \mbox{+1.12/-1.44} LSB$_{12}$ and the DNL is  \mbox{+1.33, -0.62} LSB$_{12}$. The typical size of the calibration correction for each MDAC is $2 - 7$\%. Figure \ref{fig:full_fft} shows the output FFT of a 10~MHz input sine-wave after digital calibration. The ENOB at 10~MHz for the calibrated channel is 11.0-bits, calculated from the signal to noise and distortion ratio (SINAD) of 67.9 dB, and the SFDR is 79.4~dB. The measurements are repeated at input signal frequencies of 200 kHz and 18 MHz, verifying the expected performance observed in \cite{nevis10}. The total analog power consumption of the chip is 50~mW per channel.

The dynamic performance can be compared to the expected $\sim 11.5$ ENOB performance of the MDACs \cite{nevis10}. The decrease is due to two challenges encountered with the design of the chip. The first is related to the architecture of the chip. The four MDACs in pipeline mode operating at 40~MHz are followed by the SAR operating at 640~MHz (this frequency choice was influenced by the low-jitter 640~MHz clock required for serialization). In order to fit the nine 640~MHz clock cycles (eight for the SAR operations plus one for the common mode voltage change) into exactly one half of a 40~MHz cycle both edges of the 640~MHz clock are used. This requires a well-defined duty cycle for the 640~MHz clock with respect to the 40~MHz clock. This chip does not have reconditioning hardware to restore the duty cycle during operation. This causes the MDACs and SAR to no longer be perfectly in phase. The results are spurious codes where the MDAC and SAR output are not synchronized, degrading the performance. This was confirmed with transistor-level simulations. Clock adjustments were made for individual chips which reduced the number of spurious codes. In the next generation of this chip additional hardware has been proposed to recondition the fast clock duty cycle, as well as hardware to restore the proper relationship between the slow and fast clocks at the point of use. Alternatively, a system using only one edge of the fast clock could be used.  

The second challenge was connected to the design of reference voltage drivers. Three reference voltages are necessary ($V_{ref}, V_{ref}/4, V_{ref}^{SAR}$). In one case ($V_{ref}/4$) the open-loop gain of the amplifiers controlling the switched drivers was insufficient. The resulting voltage source did not supply enough charge to its load. This was confirmed in transistor-level simulations.  An external reference voltage was provided during the testing, however the option to completely disable the internal reference voltage was not included in the design. A solution has been proposed which corrects the open-loop gain, and may be implemented in the next generation of this chip. 
 
 \subsection{Irradiation}

\begin{figure}
	\begin{center}
		\includegraphics[scale=0.2]{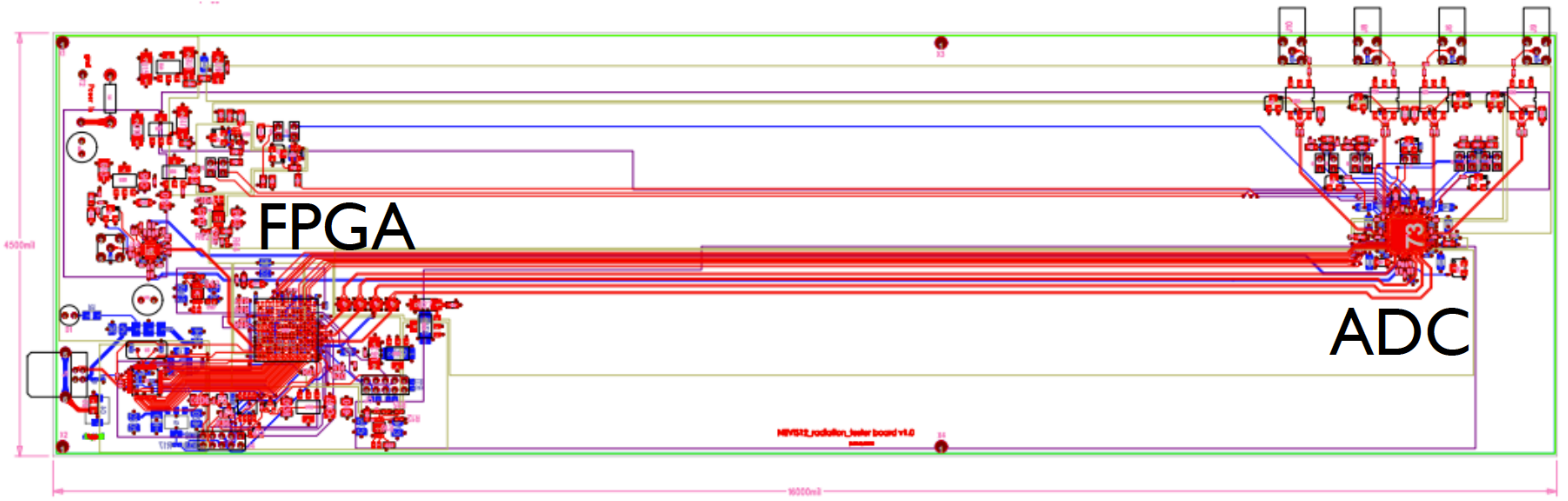}
		\caption{Test board for irradiation. The length of the board is $\sim 30$~cm in order to separate the active elements from the ADC under test. The active elements are protected by lead shielding during testing.}
		\label{fig:rad_board}
	\end{center}
\end{figure}

The goals of the irradiation testing were to measure the SEE cross section and to test the radiation tolerance of the SAR ADC (the insensitivity to  large TID of the high precision, MDAC stages was measured in~\cite{nevis10}).  Five chips were taken to the Massachusetts General Hospital Francis H. Burr Proton Therapy Center. These chips were selected using a socketed board to verify their functionality, and were then mounted on a specialized board designed to allow the readout of the chip during the irradiation. As shown in Figure \ref{fig:rad_board}, the chip was placed at one end of the $\sim 30$ cm long board, while the other active elements, including the FGPA, were placed at the other end. The high radiation flux in the ATLAS detector was replicated by irradiating four of the chips using the 227 MeV proton beam. During the irradiation, the active elements  not under test were protected with lead shielding. This board was powered and the current drawn by the chip was monitored.  A clock signal was provided, and a pure sine-wave signal was applied to the input of the ADC. The board limited the precision of the measurements to approximately 10 ENOB.  The limitation prevented the testing of the full 12-bit resolution. However, previous radiation testing of the MDAC stages established that doses as large as 10~MRad did not measurably degrade the MDAC precision \cite{nevis10}.

\begin{figure}
	\begin{center}
		\includegraphics[scale=0.50]{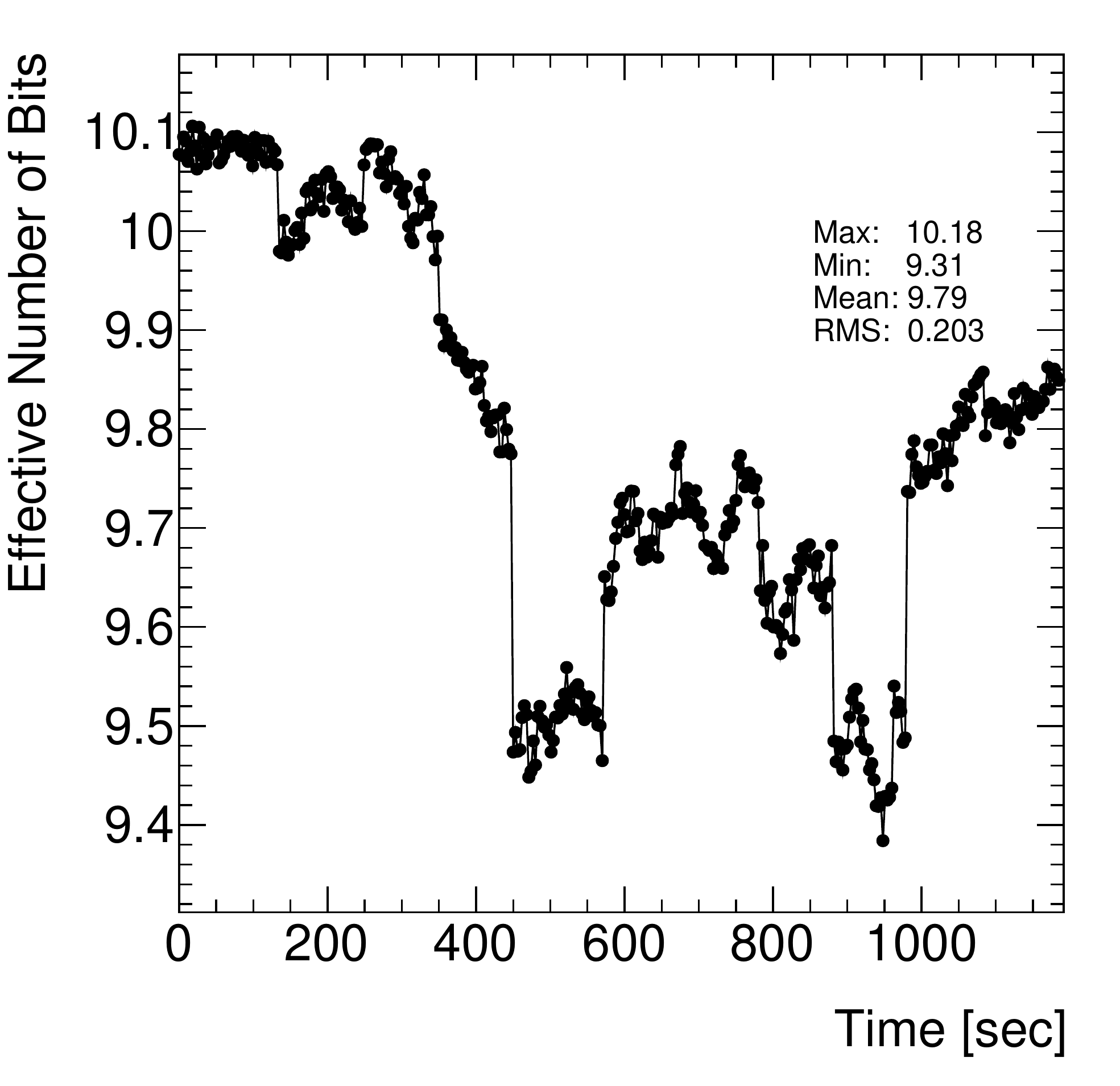}
		\caption{ENOB monitored during irradiation. The fluxuations in the ENOB are expected and caused by rerunning the calibration procedure several times during the irradiation, as would be done in the use of the device at the ATLAS experiment. Note that on the specialized irradiation board the ENOB is limited to approximately 10. 1100~s corresponds to a dose of 2~MRad, or 75~kRad (the TID specification) in less than 50~s.}
		\label{fig:irr_current}
	\end{center}
\end{figure}

The performance of the chip over a large TID was measured by irradiating two chips, one to 1~MRad and the other to 2~MRad. During the irradiation, the ENOB of two of the chips was measured every five seconds, while the calibration was performed every 120 seconds. The limited precision was sufficient to determine the 8-bit SAR ADC sensitivity to TID and the SEE cross section. The limited precision also enters the measurement through the calibration. Since the calibration constants are measured with this board, they have a reduced precision as well. This resulted in fluctuations in the ENOB after each new calibration was applied, as observed in Figure~\ref{fig:irr_current}. The test demonstrates that the ADC is functioning well after irradiation with up to 2~MRad with no sign of failures and or large degradation of ENOB performance: in future testing it may be possible to improve the precision of the test board. Table \ref{tab:irr} lists the dose for the two chips and the performance measurements before the start of the irradiation and immediately afterwards, confirming the insensitivity of the 8-bit SAR ADC to TID. The calibration constants varied by $1 - 2$\%. 

Two additional chips were irradiated to measure the SEE cross section. In this test the sine-wave signal was removed. Outputs within a window around the pedestal value were removed by a filter applied in the FPGA. In this way single-event effects could be observed as the appearance of a digital output which was outside the filter window, with very little dead time. Additionally, spurious codes (caused by the clock duty cycle issue) which had been observed before irradiation were masked offline.

Table \ref{tab:irrsee} lists the dose and cross section measurements for the two chips. There are three types of SEE observed. The first is a single-event-function-interrupt (SEFI), in which the ADC ceases to operate normally. When detected (constant ADC output is observed) an external reset returns the ADC to normal operation. This is not equivalent to a latchup: restoring operation after a latchup would require power cycling the ADC, therefore it can be concluded that latchup events were not observed. The next is a single-event-upset (SEU), which results in a transient, erroneous code. This can affect either the analog or digital portion of the chip. When the analog signal path is affected the errors can be observed as a distribution around the pedestal value, effectively reducing the resolution of the ADC. As a result of the binary scaling of the capacitors in the SAR, the sensitivity to upsets drops quickly with bit significance. That is, the SEU cross section, in this case, decreases as the magnitude of the induced error in the number of ADC counts increases. If the errors result from the digital portion of the chip (bit-flips) then the errors are typically large deviations from the pedestal value.  The total SEE is the sum of these errors. In determining the SEE cross section we exclude the analog errors, as they are primarily a resolution effect, but also indicate the value obtained if such errors were to be included.

\begin{table}
  \begin{center}
    \caption{Measurements of ADC performance before and  immediately after irradiation in a 227 MeV proton beam at $f_{in}$~=~10 MHz. The change in the ENOB is within the measurement errors of the testing setup.}
    \resizebox{\columnwidth}{!}{
    \begin{tabular}{c c c c c c}
      \hline \hline
      Chip No. &   Dose & SNDR [dBc]  & SFDR [dBc] & ENOB \\
                       & [MRad] &  Pre/Post-Irradiation &  Pre/Post-Irradiation &   Pre/Post-Irradiation\\
	\hline	
	1 & 2 & 62.43 / 61.05 & 67.27 / 70.06 & 10.08 / 9.85\\
	2 & 1 & 63.54 / 62.36 & 70.92 / 72.98 & 10.26 / 10.10\\
      \hline \hline
    \end{tabular}
    }
  \label{tab:irr}
  \end{center}
\end{table}

\begin{table}
  \begin{center}
    \caption{Measurements of ADC SEE performance in a 227 MeV proton beam.}
        \resizebox{\columnwidth}{!}{
    \begin{tabular}{c c c c c c c c}
      \hline \hline
      Chip No. & Rate & Dose & SEFI & SEU  & SEU & SEE &Cross section (w/analog errors) \\
                      &  [$10^8\mathrm{proton}/\mathrm{cm}^{2}/\mathrm{s}$] & [kRad] & & (analog) & (digital)  & &  [$10^{-12}$ cm$^2$] \\
	\hline	
	3 & 19.0 & 101 & 0 & 8 & 1 & 9     & 0.6  ($5.7\pm1.9$) \\
	3 & 76.0 & 283 & 0 & 41  & 2 & 43  & 0.6 ($9.8\pm1.5$)\\
	4 & 18.6 & 203 & 1 & 10  & 0 & 11  & 0.3 ($3.5\pm1.1$)\\
      \hline \hline
    \end{tabular}
    }
  \label{tab:irrsee}
  \end{center}
\end{table}

\section{Conclusion} \label{sec:Conclusion}
The design of a radiation-hard dual-channel 12-bit 40 MS/s Pipeline ADC test chip was presented, an important step in the development of an ADC for use in the readout electronics upgrade for the ATLAS Liquid Argon Calorimeters at the CERN LHC. The ADC was confirmed to meet the analog-to-digital conversion performance criteria with 11.0 ENOB and 67.9~dB~SNDR at 10~MHz at 40~MS/s, with a latency of 87.5~ns (to first bit read out), while its total power consumption is 50~mW/channel. The new SAR ADC was tested and found to be robust to total dose up to 2~MRad. The SEE cross section was measured to be less than $10^{-12}$ cm$^2$. Several challenges were encountered, specifically the design of the clock conditioning and the reference voltage drivers. In future designs a PLL may be used to ensure the reliable transmission of serialized output, while for the reference voltage a solution correcting the open-loop gain has been investigated. 

\section{Acknowledgments} \label{sec:Acknowledgments}
We would like to acknowledge the excellent work and essential contributions of the technical staff at Nevis Laboratory. We also thank the CERN microelectronics group for the building blocks they provided. This work has been supported by the US National Science Foundation, award number 1067934 (Columbia).



%% \bibliographystyle{elsarticle-harv} 
%% \bibliography{nevis12_NIM}


\end{document}